\documentclass[a4paper]{jpconf}
\usepackage{graphicx}
\usepackage{multicol}
\begin{document}
\title{Cu-NMR study on the disordered quantum spin magnet 
with the Bose-glass ground state
}
\author{T. Fujiwara$^1$, H. Inoue$^1$, A. Oosawa$^1$, R. Tsunoda$^1$, T. Goto$^{1,4}$, \\
T. Suzuki$^{1,3}$,
Y. Shindo$^3$, H. Tanaka$^3$,\\
T. Sasaki$^4$, N. Kobayashi$^4$, S. Awaji$^4$, K. Watanabe$^4$}

\address{$^1$Department of Physics, Sophia University, 7-1 Kioi-cho, Chiyoda-ku, 
  Tokyo 102-8554, Japan}
\address{$^2$Faculty of Science, Tokyo Institute for Technology 
  2-12-1 Ookayama, Meguro-ku, Tokyo, 152-8551, Japan}
\address{$^3$Advanced Meson Science Laboratory, RIKEN, 2-1 Hirosawa, Wako, 
  Saitama 351-0198, Japan}
\address{$^4$Institute for Material Research, Tohoku University
  2-1-1 Katahira, Aoba-ku, Sendai 980, Japan}

\ead{gotoo-t@sophia.ac.jp}

\begin{abstract}
Cu-NMR study has been performed on the disordered spin-gap system 
${\rm Tl}_{1-x}{\rm K}_x{\rm CuCl}_3$ 
In the high-field $H>H_{\rm C}$=$\Delta/\mu_{\rm B}$, where $\Delta$ is the spin-gap,
the hyperfine field becomes extremely inhomogeneous
at low temperatures due to the field-induced magnetic order, 
indicating that the ordered spin state must be different from the pure TlCuCl$_3$.
In the low field $H<H_{\rm C}$, a saturating behavior in the longitudinal nuclear 
spin relaxation rate $T_1^{-1}$ was observed at low temperatures, 
indicating existence of the magnetic ground state proposed to be
Bose-glass phase by Fisher.
\end{abstract}

   Spin systems with a low dimensionality, a small spin number, and a small 
anisotropy often show a gapped ground state by forming singlet dimers.   
The magnetic field which exceeds $H_{\rm C}$ induces triplons,
which act as nearly free bosons, and are expected 
to undergo the Bose-Einstein condensation (BEC) at low temperatures~\cite{Nikuni_BEC_1st}.  
TlCuCl$_3$ and KCuCl$_3$ are typical examples for the quantum spin magnet\cite{Tl_uniform_mag_BEC}.  
The field-induced magnetic order in these systems
has been observed experimentally,~\cite{Tl_uniform_mag_BEC} 
and interpreted theoretically as the Bose-Einstein condensation\cite{Nikuni_BEC_1st}.
The spin state in the BEC phase is considered to be the superposition of singlet 
and triplet states.~\cite{Matsumoto_dispersion_and_staggered_Tl}   
This coherent hybridization brings about two effects to the spin system, 
the increase in the uniform magnetization, and the formation 
of the staggered magnetization directed perpendicular to the applied field.   
Both the two are detected experimentally 
below the ordering temperature $T_{\rm N}(H)$.~\cite{Tl_uniform_mag_BEC,Tanaka_neutron_m_staggered}   

The motivation of this study is rooted in the interest 
in what is caused by introducing a disorder to these quantum spin magnets.
Fisher {\em et al.} have investigated the phase diagram of disordered Bose particle system 
theoretically to find the appearance of the new phase 
``Bose-glass'' at absolute zero.~\cite{Bose_glass_fisher} 
They propose characteristics of the Bose-glass phase as:
it locates in $H$-$T$ phase diagram 
on the zero-temperature line at low field region, adjacent to the BEC phase
at high field region,
2. the field-induced triplons are massless and localized, and 
3. the phase boundary is modified from the clean system.
One can introduce disorder to TlCuCl$_3$ very easily by substituting Tl site with K.   
Oosawa {\em et al}. have reported~\cite{Tl_uniform_mag_BEC} that the 
$M$-$H$ curve of (Tl,K)CuCl$_3$ has a finite gradient around zero field,
and that BEC takes place at high fields above $H_{\rm C}$.
Recent study of specific heat by Shindo {\em et al.} has revealed 
that $H_{\rm N}(T)$, the phase boundary of BEC is appreciably modified by disorder as predicted by 
Fisher.~\cite{Shindo_TlK}  
These results seem to be quite convincing for the existence of the Bose-glass phase, 
but, the microscopic spin state of the phase is unknown. 
The purpose of this study is to investigate microscopically on these two points by Cu-NMR.


The two single crystals of Tl$_{1-x}$K$_x$CuCl$_3$ ($x$=0.1, 0.2) were synthesized by the Bridgman 
method.\cite{Tl_uniform_mag_BEC,Shindo_TlK}.
The critical fields in the field parallel to the $b$-axis of the two samples $x$=0.1, 0.2 are 
$H_{\rm C}$=4.4 and 3.9T respectively \cite{Shindo_TlK}. 
NMR spectra are obtained by plotting the spin-echo amplitude against either 
the applied field
or the sample-rotating angle.
The spin-lattice relaxation rate $T_1^{-1}$ was determined by the saturation-recovery method~\cite{relaxation_curve_Narath}.

\begin{figure}[b]
\begin{center}
\includegraphics*[trim=0cm 19.9cm 0cm 1.0cm,clip,scale=0.8]{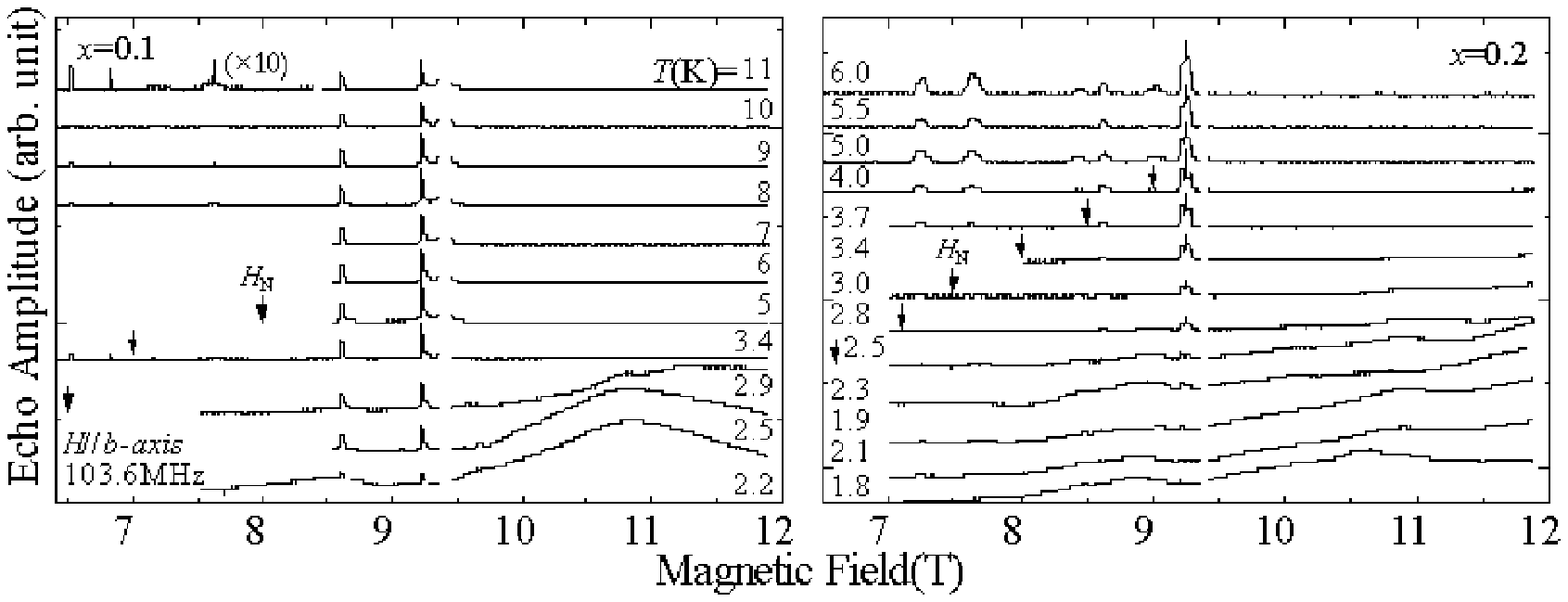}
\end{center}
\caption{Temperature dependence of field-swept NMR spectra for $x$=0.1 and 0.2
under field regions around the BEC transition.  
The position of $H_{\rm N}(T)$ at
each temperature is indicated by arrows. 
}
\end{figure}


Figure 1 shows the temperature dependence of field-swept spectra in the field region 6-12T,
where $H_{\rm N}(T)$ the transition temperature of BEC\cite{Tl_uniform_mag_BEC,Shindo_TlK} 
is shown by arrows.
At high temperatures, sharp paramagnetic NMR lines are observed, 
and explained in terms of eqQ-perturbed nuclear states for $^{63}$Cu and $^{65}$Cu 
($I$=3/2).
We show in Fig. 2 the profile of rotating spectra measured under various magnetic fields with
the peak positions calculated by the numerical diagonalization of nuclear spin hamiltonian.
The two show a good agreement with NMR parameters of 
$^{63}\nu_{\rm Q}$=39.2MHz, $\eta\simeq$0, $K\simeq$0(\%), 
and the direction of the principal axis of the electric field gradient tensor
perpendicular to the ${\rm Cu_{2}Cl_{6}}$ basal plane, 
which are the nearly same as those used to explain the singlet sites in iso-structural compounds 
KCuCl$_3$ and NH$_4$CuCl$_3$.~\cite{NH4_LT_hosoya,NH4_ICM_tani}
For there are the two dimers connected with the glide symmetry in a unit cell, 
a peak number is doubled to be twelve, though 
one of the two satellite transitions is always invisible.
The same case occurs in pure TlCuCl$_3$ but not in NH$_4$CuCl$_3$ and KCuCl$_3$.
In higher field region above 7T, the amplitude of signal in Fig. 2 becomes extremely weak
because the spin-spin relaxation time $T_2^{-1}$ tends to be very 
short due to the increase in the number of triplons.

\begin{figure}[b]
\begin{center}
\includegraphics*[trim=0cm 20.8cm 0cm 1.1cm,clip,scale=0.85]{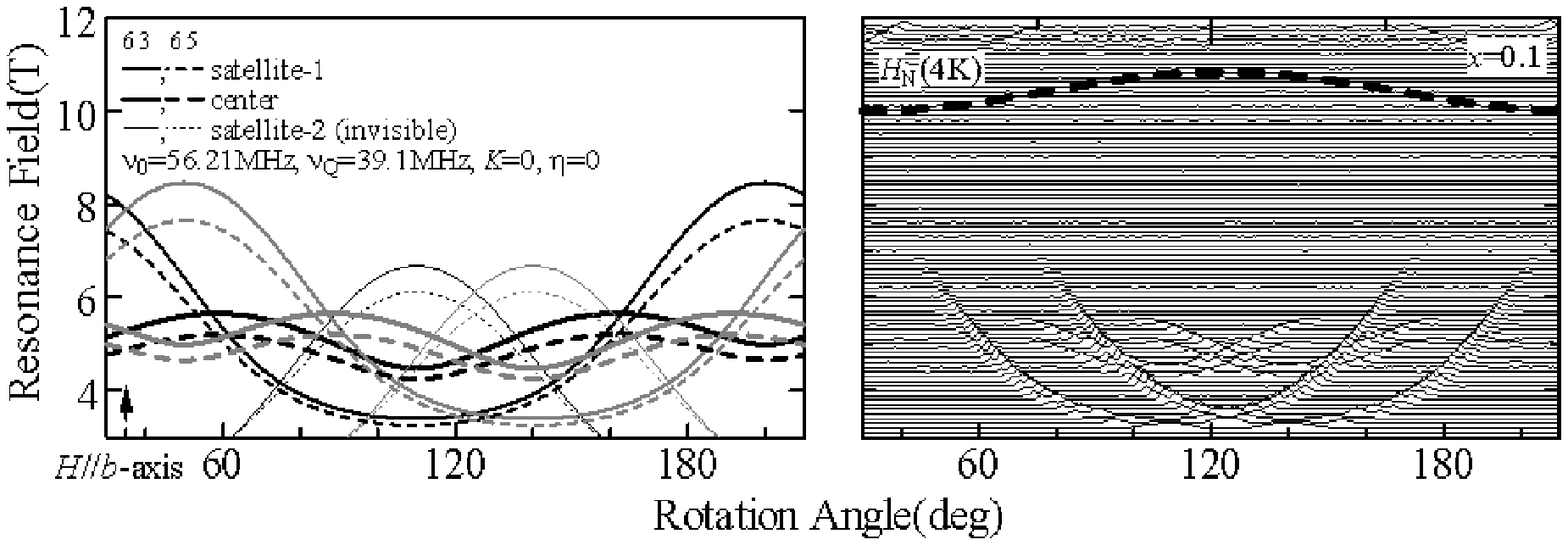}
\end{center}
\caption{Rotational profile of spectra taken under various magnetic fields
(right) and calculated peak positions
(left).  The signals that come from the two dimers in a unit cell is
shown by black and gray curves.
A dashed thick curve (right) indicates magnetic ordering field $H_{\rm N}(4K)$, 
compensated by the anisotropic $g$-tensor.\cite{Tl_uniform_mag_BEC}
}
\end{figure}

With lowering temperature, as shown in Fig. 1, 
the peak positions do not change, that is, within $\pm$0.05\%. 
in the experimental temperature range.   
When the temperature is still lowered to around 3K,
a drastic change is observed in spectra.
An extremely broad peak appears from the higher field side, and covers all the sharp peaks.  
At the lowest temperature 1.8K, the width of the broad peak exceeds 4 Tesla.
The field region of this broad tail moves to lower field
as the temperature is decreased, but the leftmost of the region seems to be always higher
than reported value of the BEC transition 
$H_{\rm N}(T)$ by 3-4T\cite{Tl_uniform_mag_BEC,Shindo_TlK}.
These behaviors of the NMR spectra is quite different from 
the pure system where the clear splitting is 
observed in the BEC state.\cite{Vyaselev_Takigawa_NMR}
Transition from the paramagnetic peaks to the broad one at 
BEC state takes place in a very narrow temperature range.
For both the samples with $x=$0.1 and 0.2, in focusing at the fixed field around 11T, 
the change occurs within $\Delta T=$0.2 K, the sharpness of which 
is comparable with that of the pure system.\cite{Vyaselev_Takigawa_NMR}

%


Figure 3 shows the temperature dependence of the relaxation rate $T_1^{-1}$ 
under various fields higher (A, B) or lower (C, D) than $H_{\rm C}$=$H_{\rm N}(0)$.
In the former case of (A), $T_1^{-1}$ first shows a significant decrease 
with decreasing temperature from the paramagnetic region, 
and then a diverging behavior at $T_{\rm N}(T)$.
In the panel (B), the diverging behavior toward $T_{\rm N}(T)$ from 
the low temperature side is shown.
The similar behavior is reported for pure TlCuCl$3$\cite{Vyaselev_Takigawa_NMR}.
In low fields shown in (C) and (D), where the ground state is expected
to be Bose-glass,
$T_1^{-1}$ decreases steeply until 4K, where it saturates and stays constant.
This behavior is completely different from what is expected in the gapped state.
The saturated value of $T_1^{-1}$ at lowest temperature depends on 
the mixing ratio $x$.   The sample $x$=0.2 shows $T_1^{-1}\simeq 1$(msec),
which is ten times larger than that of $x$=0.1.


The drastic broadening of spectra observed in BEC state
at low temperatures and high fields shown in Fig. 1 indicates
that the spin state in the BEC of the disordered system is completely different 
from that in pure TlCuCl$_3$. 
In the BEC state of the pure system, 
the appearance of the homogeneous staggered magnetization perpendicular to the applied
field has been reported by neutron and NMR~\cite{Tanaka_neutron_m_staggered,Vyaselev_Takigawa_NMR}.
Our result on the disordered system (Tl,K)CuCl$_3$ shows that
the hyperfine field differs by site by site, and is extremely inhomogeneous, 
though its averaged value is similar to the pure system.
The induced magnetization is of staggered-type, because the broad peak spreads over 
toward the both sides of the zero-shift position.   
In the vicinity of $T=T_{\rm N}$, the broad peak appears from the higher field side,
because the BEC transition temperature $T_{\rm N}$ increases monotonically with 
increasing field.\cite{Bose_glass_fisher,Tl_uniform_mag_BEC,Shindo_TlK}   

\begin{figure}[t]
\begin{center}
\includegraphics*[trim=0cm 21cm 0cm 1.9cm, clip, scale=0.85]{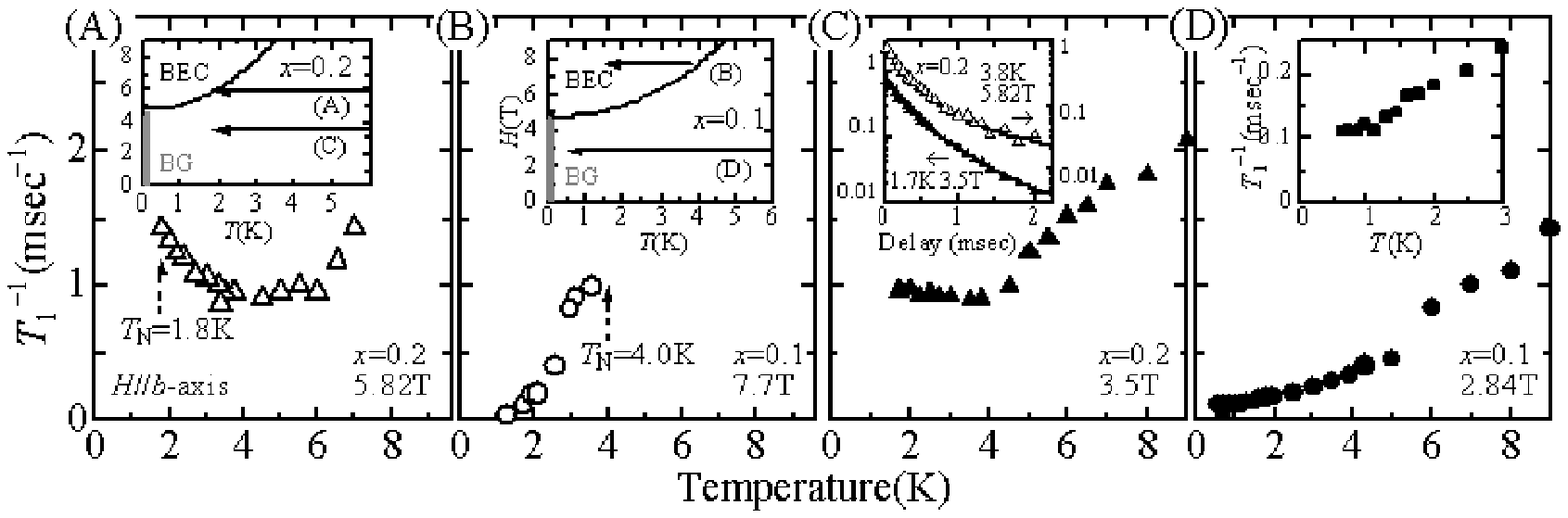}
\end{center}
\caption{The temperature dependence of $T_1^{-1}$ under various fields above
and below $H_{\rm N}(T=0)$.  
Measured fields and phase diagrams showing the area of BEC and
Bose-glass(BG) are shown in insets of A and B.
The inset in C shows typical relaxation curves for the satellite transition
and the fitting function
$0.1e^{-t/T_1}+0.5e^{-3t/T_1}+0.4e^{-6t/T_1}$.   }
\end{figure}

The origin of the inhomogeneity in the hyperfine field is explained as follows.
The Bose-Einstein condensation in TlCuCl$_3$
derives its name from the fact that the Bloch state of 
$|{\rm S}_n\rangle+A\cdot e^{i({\bf q} \cdot {\bf r}_n-\omega_q t)} |{\rm T}^+_n\rangle $
condensates at the lowest energy state with $q=Q_{\rm BEC}$, 
which is reported to be (0,0,1).\cite{Tanaka_neutron_m_staggered,Matsumoto_dispersion_and_staggered_Tl}
The magnitude and the direction of the staggered moment are
proportional to the hybridization amplitude and its phase.
The broad spectra observed in the disordered system indicate that either the amplitude
or the direction of the staggered magnetization must be distributed.
In (Tl,K)CuCl$_3$ system, the strength of the spin-spin interaction is randomly
modulated by the structural randomness due to the heteroatomic substitution.
This modulation is expected to bring the inhomogeneity 
in the hybridization amplitude or its phase.
When the direction of this staggered magnetization is not parallel with principal axes
of either the hyperfine tensor or the quadrapole tensor, it causes a differentiation
of the NMR shift in the two nuclei in the dimer, and hence the splitting or broadening
in spectra.

The temperature independent shift in the paramagnetic state
is explained in terms of the triplon localization in disordered 
systems\cite{Shindo_TlK,NH4_LT_hosoya,NH4_ICM_tani}.   
In (Tl,K)CuCl$_3$ system, induced triplon are localized in space by disorder, 
so that they contribute to macroscopic magnetization, but to the NMR shift of the singlet site.
This behavior is quite different from the case of the pure TlCuCl$_3$\cite{Vyaselev_Takigawa_NMR}, 
where the induced triplons move around freely to produce the local magnetization uniform in the entire sample.


The existence of the critical slowing down at the vicinity of $T_{\rm N}$ is consistent with
the fact that BEC is the second order phase transition.
The saturating behavior of $T_1^{-1}$ at low temperature in the low field region 
indicates that the ground state is gapless, which is consistent with the Fisher's proposal on the
Bose-glass.   Suzuki {\em et al.} recently have reported the similar behavior 
in the muon relaxation rate\cite{Suzuki_muon}.


\section*{Acknowledgments}
The authors are grateful for discussion Prof. M. Matsumoto.
A part of experiments has been performed at HFLSM, Inst. Mat. Res., Tohoku Univ.   
This work is supported by the Toray Science Foundation, Saneyoshi Foundation, 
Kurata foundation and a Grant-in-Aid for Scientific Research from the Ministry of Education, 
Culture, Sports, Science and Technology of Japan.

\section*{References}
\vspace*{-4mm}
\begin{multicols}{2}

\end{multicols}


\begin{thebibliography}{22}

\bibitem{Nikuni_BEC_1st} Nikuni T {\em et al.} 
2000 {\it Phys. Rev. Lett.} {\bf 84} 5868


\bibitem{Tl_uniform_mag_BEC} 
Oosawa A {\em et al.} 
1999 {\it J. Phys. Condens. Matter} {\bf 11} 265-271, 
2001 {\it Phys. Rev.} {\bf B63}, 134416,
2001 {\it Phys. Rev.} {\bf B63}, 134416,
2002 {\it Phys. Rev.} {\bf B65}, 094426,
2002 {\it Phys. Rev.} {\bf B65}, 184437





\bibitem{structure} Willet R D {\em et al.}
1963 {\it J. Chem. Phys.} {\bf 38}, 2429

\bibitem{inelastic_neutron_Tl}
Cavadini N {\em et al.}
2002 {\it Phys. Rev.}{\bf B65}, 132415


\bibitem{Matsumoto_dispersion_and_staggered_Tl}
Matsumoto M {\em et al.}
2002 {\it Phys. Rev. Lett.} {\bf 89}, 077203, 2004 {\it Phys. Rev.} {\bf B69}, 054423


\bibitem{Tanaka_neutron_m_staggered}
Tanaka H {\em et al.}
2001 {\it J. Phys. Soc. Jpn.} {\bf 70}, 939


\bibitem{Shindo_TlK} Shindo Y, Tanaka H
2004 {\it J. Phys. Soc. Jpn.} {\bf 73}, 2642

\bibitem{Bose_glass_fisher}
Fisher M P A {\em et al.}
1989 {\it Phys. Rev.} {\bf B40}, 546.


\bibitem{relaxation_curve_Narath}
Narath A
1967 {\it Phys. Rev.} {\bf 162}, 320



\bibitem{Shiramura_TlandK}Shiramura W {\em et al.}
1997 {\it J. Phys. Soc. Jpn.} {\bf 66} 1900, 1988 {\it ibid.} {\bf 67} 1548.




\bibitem{Vyaselev_Takigawa_NMR} Vyaselev O {\em et al.}
2004 {\it Phys. Rev. Lett.} {\bf 92}, 207202 








\bibitem{NH4_LT_hosoya}Hosoya S {\em et al.}
2003 {\it Physica} {\bf B329-333} 977.


\bibitem{NH4_ICM_tani}Tani S {\em et al.}
2004 {\it J. Magn. Magn. Mater.} {\bf 272-276} (2004)





\bibitem{Suzuki_muon}Suzuki T {\em et al.}
2006 {\it J. Phys. Soc. Jpn.} {\bf 75} 025001.


\end{thebibliography}
\end{document}